\documentclass[showkeys,showpacs,prd]{revtex4}
\usepackage{amsmath}
\usepackage{amssymb}
\usepackage{graphicx}
\usepackage{color}
\usepackage{ulem}
\usepackage{tikz}

\begin{document}

\title{
The simplest nonassociative generalization of supersymmetry
}

\author{
Vladimir Dzhunushaliev
}
\email{v.dzhunushaliev@gmail.com}
\affiliation{
Dept. Theor. and Nucl. Phys., KazNU, Almaty, 050040, Kazakhstan
}
\affiliation{
IETP, Al-Farabi KazNU, Almaty, 050040, Kazakhstan
}

\begin{abstract}
Nonassociative generalization of supersymmetry is suggested. 3- and 4-point associators for
supersymmetric generators are considered. On the basis of zero Jacobiators for three supersymmetric
generators, we have obtained the simplest form of 3-point associators. The connection between 3- and 4-point
associators is  considered. On the basis of this connection, 4-point associators are obtained.
The Jacobiators for the product of four supersymmetric generators are calculated.
We discuss the possible physical meaning of numerical coefficients presented on the right-hand sides of associators.
The possible connection between supersymmetry, hidden variables, and nonassociativity is discussed.
\end{abstract}


\maketitle

\section{Introduction}

Supersymmetry is a well-defined mathematical theory that probably has the application in physics:
it is a branch of particle physics that, using a proposed type of spacetime symmetry, relates
two basic classes of elementary particles~-- bosons and fermions. In the standard approach supersymmetric
generators are associative and anticommutative. Here we want to consider a nonassociative generalization
of supersymmetry. We offer some 3-point associators for supersymmetric generators $Q_a$ and $Q_{\dot a}$.
Using some relation between 3- and 4-point associators, we  obtain some limitations on the possible form of
4-point associators. On the basis of these limitations, we  offer 4-point associators for supersymmetric generators.

Nonassociative structures appear in: (a) quantum chromodynamics \cite{Gunaydin:1974};
(b) Maxwell and Dirac equations \cite{Gogberashvili:2005xb,Gogberashvili:2005cp}; (c)~string theory \cite{Gunaydin:2013nqa};
(d) nonassociative quantum mechanics \cite{Bojowald:2015cha,Dzhunushaliev:2005yd}.
For other ways of introducing  nonassociative structures into physics, see the monographs \cite{okubo,gursey}.

Here we would like to introduce nonassociative structures into supersymmetry and to discuss the physical consequences of such a procedure.

\section{The simplest 3-point associators}

Recall the definition of associator
\begin{equation}\label{2-10}
\left[ A, B, C \right] = \left( A B \right) C - A \left( B C \right),
\end{equation}
where $A,B,C$ are nonassociative quantities. Now we want to demonstrate that it is possible
to introduce the  generalization of supersymmetry other than that given in Refs.~\cite{Dzhunushaliev:2015fia,Dzhunushaliev:AACA}.
Let us define the following 3-point associators:
\begin{eqnarray}
	\left[
		Q_a, Q_b, Q_c
	\right] &=& \alpha_1 Q_a \epsilon_{bc} +
	\alpha_2 Q_b \epsilon_{ac} +
	\alpha_3 Q_c \epsilon_{ab} ,
\label{2-20}\\
	\left[
		Q_{\dot a}, Q_b, Q_c
	\right] &=& \beta_1 Q_{\dot a} \epsilon_{bc} ,
\label{2-30}\\
	\left[
		Q_a, Q_{\dot b}, Q_c
	\right] &=& \beta_2 Q_{\dot b} \epsilon_{ac} ,
\label{2-40}\\
	\left[
		Q_a, Q_b, Q_{\dot c}
	\right] &=& \beta_3 Q_{\dot c} \epsilon_{ab} ,
\label{2-50}\\
	\left[ Q_a, Q_{\dot b}, Q_{\dot c}
	\right] &=& \gamma_1 Q_a \epsilon_{\dot b \dot c} ,
\label{2-60}\\
	\left[ Q_{\dot a}, Q_b, Q_{\dot c}
	\right] &=& \gamma_2 Q_b \epsilon_{\dot a \dot c} ,
\label{2-70}\\
	\left[ Q_{\dot a}, Q_{\dot b}, Q_c
	\right] &=& \gamma_3 Q_c \epsilon_{\dot a \dot b} ,
\label{2-80}\\
	\left[ Q_{\dot a}, Q_{\dot b}, Q_{\dot c} \right] &=&
	\delta_1 Q_{\dot a} \epsilon_{\dot b \dot c} +
	\delta_2 Q_{\dot b} \epsilon_{\dot a \dot c} +
	\delta_3 Q_{\dot c} \epsilon_{\dot a \dot b},
\label{2-90}
\end{eqnarray}
where
\begin{eqnarray}
	\epsilon_{\dot a \dot b} = \epsilon_{a b} &=&
	\left(
		\begin{array}{cc}
		0          &     1     \\
		-1          &     0
	\end{array}
	\right),
\label{2-90a}\\
	\epsilon^{ab} = \epsilon^{\dot a \dot b} &=&
	\left(
	\begin{array}{cc}
	0          &     -1     \\
	1          &     0
	\end{array}
	\right).
\label{2-90b}
\end{eqnarray}
The dotted indices are lowered and raised by using
$\epsilon_{\dot a \dot b}, \epsilon^{\dot a \dot b}$, and the undotted -- by
$\epsilon_{a b}, \epsilon^{ab}$.

In order to introduce some limitations on the nonassociative algebra, we want to calculate the Jacobiator
\begin{equation}\label{2-100}
	J(x, y, z) = \left[
	\left[
	x, y
	\right] , z
	\right] +
	\left[
	\left[
	y, z
	\right] , x
	\right] +
	\left[
	\left[
	z, x
	\right] , y
	\right] = \left[ x, y, z \right] + \left[ y, z, x \right] +
	\left[ z, x, y \right] -
	\left[ x, z, y \right] - \left[ y, x, z \right] -
	\left[ z, y, x \right],
\end{equation}
where $x,y,z$ are either $Q_{a, \dot a}$ or their product. Let us calculate Jacobiators
\begin{eqnarray}
	J(Q_a, Q_b, Q_c) &=& 2 \left(
		\alpha_1 - \alpha_2 + \alpha_3
	\right) Q_a \epsilon_{bc} ,
\label{2-110}\\
	J(Q_{\dot a}, Q_b, Q_c) &=& 2 \left(
	\beta_1 - \beta_2 + \beta_3
	\right) Q_{\dot a} \epsilon_{bc} ,
\label{2-120}\\
	J(Q_a, Q_{\dot b}, Q_c) &=& 2 \left(
	\beta_1 - \beta_2 + \beta_3
	\right) Q_{\dot b} \epsilon_{ca} ,
\label{2-130}\\
	J(Q_a, Q_b, Q_{\dot c}) &=& 2 \left(
	\beta_1 - \beta_2 + \beta_3
	\right) Q_{\dot c} \epsilon_{ab} ,
\label{2-140}\\
	J(Q_a, Q_{\dot b}, Q_{\dot c}) &=& 2 \left(
	\gamma_1 - \gamma_2 + \gamma_3
	\right) Q_a \epsilon_{\dot b \dot c} ,
\label{2-150}\\
	J(Q_{\dot a}, Q_b, Q_{\dot c}) &=& 2 \left(
	\gamma_1 - \gamma_2 + \gamma_3
	\right) Q_b \epsilon_{\dot c \dot a} ,
\label{2-160}\\
	J(Q_{\dot a}, Q_{\dot b}, Q_c) &=& 2 \left(
	\gamma_1 - \gamma_2 + \gamma_3
	\right) Q_c \epsilon_{\dot a \dot b} ,
\label{2-170}\\
	J(Q_{\dot a}, Q_{\dot b}, Q_{\dot c}) &=& 2 \left(
	\delta_1 - \delta_2 + \delta_3
	\right) Q_{\dot a} \epsilon_{\dot b \dot c} .
\label{2-180}
\end{eqnarray}
To have zero Jacobiators, we have to have the following equations for the parameters $\alpha, \beta, \gamma, \delta$:
\begin{equation}
	\alpha_1 - \alpha_2 + \alpha_3 =
	\beta_1 - \beta_2 + \beta_3 =
	\gamma_1 - \gamma_2 + \gamma_3 =
	\delta_1 - \delta_2 + \delta_3 = 0.
\label{2-180a}
\end{equation}
Thus, we see that perhaps the most natural and the simplest choice of the parameters $\alpha, \beta, \gamma, \delta$ is
\begin{eqnarray}
	\alpha_2 &=& \beta_2 = \gamma_2 = \delta_2 = 0 ,
\label{2-185}\\
	\left| \alpha_1 \right| &=&
	\left| \alpha_3 \right| =
	\left| \beta_1 \right| =
	\left| \beta_3 \right| =
	\left| \gamma_1 \right| =
	\left| \gamma_3 \right| =
	\left| \delta_1 \right| =
	\left| \delta_3 \right| = \frac{\hbar}{\ell_0},
\label{2-220}
\end{eqnarray}
where the factor $\hbar/\ell_0$ is introduced for equalizing the dimensionality of the
left-hand sides and right-hand sides of equations \eqref{2-20}-\eqref{2-90}; $\ell_0$ is some characteristic length. Thus
\begin{eqnarray}
	\alpha_1 = - \alpha_3  &=& \frac{\hbar}{\ell_0} \zeta_1 ,
\label{2-260a}\\
	\beta_1 = - \beta_3 &=& \frac{\hbar}{\ell_0} \zeta_2 ,
\label{2-260b}\\
	\gamma_1 = - \gamma_3 &=& \frac{\hbar}{\ell_0} \zeta_3 ,
\label{2-260c}\\
	\delta_1 = - \delta_3 &=& \frac{\hbar}{\ell_0} \zeta_4,
\label{2-260d}\\
\end{eqnarray}
where
\begin{equation}
	\zeta_{1,2,3,4} = \begin{cases}
	\text{either} 	& \pm 1 \\
	\text{or} 		& \pm i
	\end{cases} .
\label{2-265}
\end{equation}
Thus we have the following 3-point associators
\begin{eqnarray}
\left[
	Q_a, Q_b, Q_c
		\right] &=& \frac{\hbar}{\ell_0} \zeta_1
	\left( Q_a \epsilon_{bc} - Q_c \epsilon_{ab}
	\right) ,
\label{2-70a}\\
	\left[ Q_{\dot a}, Q_b, Q_c	\right] &=&
	\frac{\hbar}{\ell_0} \zeta_2 Q_{\dot a} \epsilon_{bc} ,
\label{2-70b}\\
	\left[
	Q_a, Q_{\dot b}, Q_c
	\right] &=& 0 ,
\label{2-70c}\\
	\left[ Q_a, Q_b, Q_{\dot c} \right] &=&
	- \frac{\hbar}{\ell_0} \zeta_2 Q_{\dot c} \epsilon_{ab} ,
\label{2-70d}\\
	\left[ Q_a, Q_{\dot b}, Q_{\dot c} \right] &=&
	\frac{\hbar}{\ell_0} \zeta_3 Q_a \epsilon_{\dot b \dot c} ,
\label{2-70e}\\
	\left[ Q_{\dot a}, Q_b, Q_{\dot c} \right] &=& 0 ,
\label{2-70f}\\
	\left[ Q_{\dot a}, Q_{\dot b}, Q_c \right] &=&
	- \frac{\hbar}{\ell_0} \zeta_3 Q_c \epsilon_{\dot a \dot b} ,
\label{2-70g}\\
	\left[ Q_{\dot a}, Q_{\dot b}, Q_{\dot c} \right] &=&
	\frac{\hbar}{\ell_0} \zeta_4 \left(
		Q_{\dot a} \epsilon_{\dot b \dot c} -
		Q_{\dot c} \epsilon_{\dot a \dot b}
	\right).
\label{2-70h}
\end{eqnarray}

\section{4-point associators}

First, we want to consider the connection between 3- and 4-point associators. For example, the 4-point associator $[Q_x Q_y, Q_z, Q_w]$ is
\begin{equation}
	\left[ Q_x Q_y, Q_z, Q_w \right]  =
	\left(
	\left(
	Q_x Q_y
	\right) Q_z
	\right) Q_w -
	\left(
	Q_x Q_y
	\right)
	\left(
	Q_z Q_w
	\right),
\label{3-10}
\end{equation}
where $x,y,z,w$ are any combinations of dotted and undotted indices. The last term on the right-hand side of equation \eqref{3-10} is
$
	\left(
	Q_x Q_y
	\right)
	\left(
	Q_z Q_w
	\right)
$ and it cannot be obtained by multiplying of any 3-point associator from the left-hand sides of \eqref{2-20}-\eqref{2-90} neither by $Q_a$ nor $Q_{\dot a}$.
Nevertheless, there is the relation between the 3- and 4-point associators:
\begin{equation}
	\left[ Q_x Q_y, Q_z, Q_w \right]  -
	\left[ Q_x, Q_y Q_z, Q_w \right]  +
	\left[ Q_x, Q_y, Q_z Q_w \right]  =
	\left[ Q_x, Q_y, Q_z \right]  Q_w -
	Q_x \left[ Q_y, Q_z, Q_w \right] .
\label{3-20}
\end{equation}

\subsection{4-point associators without dots}
\label{without}

Let us assume the following 4-point associators
\begin{eqnarray}
	\left[ Q_a Q_b, Q_c, Q_d \right]  &=& \rho_{1,1} Q_a Q_b \epsilon_{cd} +
	\rho_{2,1} Q_a Q_c \epsilon_{bd} +
	\rho_{3,1} Q_a Q_d \epsilon_{bc} +
	\rho_{4,1} Q_b Q_c \epsilon_{ad} +
	\rho_{5,1} Q_b Q_d \epsilon_{ac} +
	\rho_{6,1} Q_c Q_d \epsilon_{ab} ,
\label{3-30}\\
	\left[  Q_a, Q_b Q_c, Q_d \right]  &=& \mu_{1, 1} Q_a Q_b \epsilon_{cd} +
	\mu_{2,1} Q_a Q_c \epsilon_{bd} +
	\mu_{3,1} Q_a Q_d \epsilon_{bc} +
	\mu_{4,1} Q_b Q_c \epsilon_{ad} +
	\mu_{5,1} Q_b Q_d \epsilon_{ac} +
	\mu_{6,1} Q_c Q_d \epsilon_{ab} ,
\label{3-40}\\
	\left[  Q_a, Q_b, Q_c Q_d \right]  &=& \nu_{1, 1} Q_a Q_b \epsilon_{cd} +
	\nu_{2,1} Q_a Q_c \epsilon_{bd} +
	\nu_{3,1} Q_a Q_d \epsilon_{bc} +
	\nu_{4,1} Q_b Q_c \epsilon_{ad} +
	\nu_{5,1} Q_b Q_d \epsilon_{ac} +
	\nu_{6,1} Q_c Q_d \epsilon_{ab} .
\label{3-50}
\end{eqnarray}
Then the relation \eqref{3-20} gives us the following relations between 3- and 4-point nonassociative  structure constants
\begin{eqnarray}
	\rho_{1,1} - \mu_{1,1} + \nu_{1,1} &=& \alpha_{1} ,
\label{3-60}\\
	\rho_{2,1} - \mu_{2,1} + \nu_{2,1} &=& \alpha_{2} ,
\label{3-70}\\
	\rho_{3,1} - \mu_{3,1} + \nu_{3,1} &=& \alpha_{1} + \alpha_{3},
\label{3-80}\\
	\rho_{4,1} - \mu_{4,1} + \nu_{4,1} &=& 0 ,
\label{3-90}\\
	\rho_{5,1} - \mu_{5,1} + \nu_{5,1} &=& \alpha_{2} ,
\label{3-100}\\
	\rho_{6,1} - \mu_{6,1} + \nu_{6,1} &=& \alpha_{3} .
\label{3-110}
\end{eqnarray}
Perhaps  the simplest limitations on the nonassociative structure constants $\rho_{i,1}, \mu_{i,1}, \nu_{i,1}$ are as follows:
\begin{eqnarray}
	\rho_{2,1} &=& \rho_{5,1}, \mu_{2,1} = \mu_{5 1}, \nu_{2,1} = \nu_{5,1} ;
\label{3-120}\\
	\rho_{3,1} &=& \rho_{1,1} + \rho_{6,1},
	\mu_{3,1} = \mu_{1,1} + \mu_{6,1},
	\nu_{3,1} = \nu_{1,1} + \nu_{6,1} ;
\label{3-130}\\
	\rho_{4,1} - \mu_{4,1} + \nu_{4,1} &=& 0 .
\label{3-140}
\end{eqnarray}
Then the following solution of equations \eqref{3-120}-\eqref{3-140} that is compatible with \eqref{2-185} and \eqref{2-260a} can be found:
\begin{eqnarray}
	\rho_{1,1} &=& - \rho_{6,1} = \nu_{1,1} = - \nu_{6,1} =
	\frac{1}{2} \frac{\hbar}{\ell_0} \zeta_1 ;
\label{3-150}\\
	\mu_{i,1} &=& 0, i = 1,2 , \ldots 6 ,
\label{3-160}\\
	\rho_{2,1} &=& \rho_{3,1} = \rho_{4,1} = \rho_{5,1} =
	\nu_{2,1} = \nu_{3,1} = \nu_{4,1} = \nu_{5,1} = 0 .
\label{3-180}
\end{eqnarray}
Finally, 4-point associators are
\begin{eqnarray}
	\left[ Q_a Q_b, Q_c, Q_d \right]  &=&
	\frac{1}{2} \frac{\hbar}{\ell_0} \zeta_1
	\left(
		Q_a Q_b \epsilon_{cd} -
		 Q_c Q_d \epsilon_{ab}
	\right) ,
\label{3-190}\\
	\left[  Q_a, Q_b Q_c, Q_d \right]  &=& 0 ,
\label{3-200}\\
	\left[  Q_a, Q_b, Q_c Q_d \right]  &=&
	\frac{1}{2} \frac{\hbar}{\ell_0} \zeta_1
	\left(
		Q_a Q_b \epsilon_{cd} -
		Q_c Q_d \epsilon_{ab}
	\right) .
\label{3-210}
\end{eqnarray}
One can immediately check that the Jacobiator
\begin{equation}
	J\left( Q_a Q_b, Q_c, Q_d \right) = 0.
\label{3-220}
\end{equation}	

\subsection{4-point associators with one dot}
\label{3b}

In this section we consider 4-point associators with one dot moving from the left on the right side of associator.

\subsubsection{1-st case}

We seek 4-point associators with one dot as follows:
\begin{eqnarray}
	\left[ Q_{\dot a} Q_b, Q_c, Q_d \right]  &=&
	\rho_{1,2} \epsilon_{cd} Q_{\dot a} Q_b +
  \rho_{2,2} \epsilon_{bd} Q_{\dot a} Q_c +
  \rho_{3,2} \epsilon_{b c} Q_{\dot a} Q_d +
\nonumber \\
  &&
  \rho_{4,2} \left( Q_b Q_c, x_{d \dot a} \right) +
  \rho_{5,2} \left( Q_b Q_d, x_{c \dot a} \right) +
  \rho_{6,2} \left( Q_c Q_d, x_{b \dot a} \right) ,
\label{3b-1}\\
	\left[  Q_{\dot a}, Q_b Q_c, Q_d \right]  &=&
  \mu_{1,2} \epsilon_{cd} Q_{\dot a} Q_b +
  \mu_{2,2} \epsilon_{bd} Q_{\dot a} Q_c +
  \mu_{3,2} \epsilon_{b c} Q_{\dot a} Q_d +
\nonumber \\
  &&
  \mu_{4,2} \left( Q_b Q_c, x_{d \dot a} \right) +
  \mu_{5,2} \left( Q_b Q_d, x_{c \dot a} \right) +
  \mu_{6,2} \left( Q_c Q_d, x_{b \dot a} \right) ,
\label{3b-2}\\
	\left[  Q_{\dot a}, Q_b, Q_c Q_d \right]  &=&
	\nu_{1,2} \epsilon_{cd} Q_{\dot a} Q_b +
  \nu_{2,2} \epsilon_{bd} Q_{\dot a} Q_c +
  \nu_{3,2} \epsilon_{b c} Q_{\dot a} Q_d +
\nonumber \\
  &&
  \nu_{4,2} \left( Q_b Q_c, x_{d \dot a} \right) +
  \nu_{5,2} \left( Q_b Q_d, x_{c \dot a} \right) +
  \nu_{6,2} \left( Q_c Q_d, x_{b \dot a} \right) .
\label{3b-3}
\end{eqnarray}

\subsubsection{2-nd case}

We seek 4-point associators with one dot as follows:
\begin{eqnarray}
	\left[  Q_a Q_{\dot b}, Q_c, Q_d \right]  &=&
	\rho_{1,3} \epsilon_{c d} Q_a Q_{\dot b} +
  \rho_{2,3} \left( Q_a Q_{c}, x_{d \dot b} \right) +
  \rho_{3,3} \left( Q_a Q_{d}, x_{c \dot b} \right) +
\nonumber \\
  &&
  \rho_{4,3} \epsilon_{a d} Q_{\dot b} Q_c +
  \rho_{5,3} \epsilon_{a c} Q_{\dot b} Q_d +
  \rho_{6,3} \left( Q_c Q_{d}, x_{a \dot b} \right) ,
\label{3b-4}\\
	\left[  Q_a, Q_{\dot b} Q_c, Q_d \right]  &=&
  \mu_{1,3} \epsilon_{c d} Q_a Q_{\dot b} +
  \mu_{2,3} \left( Q_a Q_{c}, x_{d \dot b} \right) +
  \mu_{3,3} \left( Q_a Q_{d}, x_{c \dot b} \right) +
\nonumber \\
  &&
  \mu_{4,3} \epsilon_{a d} Q_{\dot b} Q_c +
  \mu_{5,3} \epsilon_{a c} Q_{\dot b} Q_d +
  \mu_{6,3} \left( Q_c Q_{d}, x_{a \dot b} \right) ,
\label{3b-5}\\
	\left[  Q_a, Q_{\dot b}, Q_c Q_d \right]  &=&
	\nu_{1,3} \epsilon_{c d} Q_a Q_{\dot b} +
  \nu_{2,3} \left( Q_a Q_{c}, x_{d \dot b} \right) +
  \nu_{3,3} \left( Q_a Q_{d}, x_{c \dot b} \right) +
\nonumber \\
  &&
  \nu_{4,3} \epsilon_{a d} Q_{\dot b} Q_c +
  \nu_{5,3} \epsilon_{a c} Q_{\dot b} Q_d +
  \nu_{6,3} \left( Q_c Q_{d}, x_{a \dot b} \right) .
\label{3b-6}
\end{eqnarray}

\subsubsection{3-rd case}

We seek 4-point associators with one dot as follows:
\begin{eqnarray}
	\left[ Q_a Q_b, Q_{\dot c}, Q_d \right]  &=&
  \rho_{1,4} \left( Q_a Q_{b}, x_{d \dot c} \right) +
  \rho_{2,4} \epsilon_{b d} Q_{a} Q_{\dot c} +
  \rho_{3,4} \left( Q_a Q_{d}, x_{b \dot c} \right) +
\nonumber \\
  &&
  \rho_{4,4} \epsilon_{a d} Q_{b} Q_{\dot c} +
  \rho_{5,4} \left( Q_b Q_{d}, x_{a \dot c} \right) +
  \rho_{6,4} \epsilon_{a b} Q_{\dot c} Q_{d} ,
\label{3b-7}\\
	\left[  Q_a, Q_b Q_{\dot c}, Q_d \right]  &=&
  \mu_{1,4} \left( Q_a Q_{b}, x_{d \dot c} \right) +
  \mu_{2,4} \epsilon_{b d} Q_{a} Q_{\dot c} +
  \mu_{3,4} \left( Q_a Q_{d}, x_{b \dot c} \right) +
\nonumber \\
  &&
  \mu_{4,4} \epsilon_{a d} Q_{b} Q_{\dot c} +
  \mu_{5,4} \left( Q_b Q_{d}, x_{a \dot c} \right) +
  \mu_{6,4} \epsilon_{a b} Q_{\dot c} Q_{d} ,
\label{3b-8}\\
	\left[ Q_a, Q_b, Q_{\dot c} Q_d \right]  &=&
  \nu_{1,4} \left( Q_a Q_{b}, x_{d \dot c} \right) +
  \nu_{2,4} \epsilon_{b d} Q_{a} Q_{\dot c} +
  \nu_{3,4} \left( Q_a Q_{d}, x_{b \dot c} \right) +
\nonumber \\
  &&
  \nu_{4,4} \epsilon_{a d} Q_{b} Q_{\dot c} +
  \nu_{5,4} \left( Q_b Q_{d}, x_{a \dot c} \right) +
  \nu_{6,4} \epsilon_{a b} Q_{\dot c} Q_{d} .
\label{3b-9}
\end{eqnarray}

\subsubsection{4-th case}

We seek 4-point associators with one dot as follows:
\begin{eqnarray}
	\left[ Q_a Q_b, Q_c Q_{\dot d} \right]  &=&
  \rho_{1,5} \left( Q_a Q_b, x_{c \dot d} \right) +
  \rho_{2,5} \left( Q_a Q_c, x_{b \dot d} \right) +
  \rho_{3,5} \epsilon_{b c} Q_a Q_{\dot d} +
\nonumber \\
  &&
  \rho_{4,5} \left( Q_b Q_c, x_{a \dot d} \right) +
  \rho_{5,5} \epsilon_{a c} Q_b Q_{\dot d} +
  \rho_{6,5} \epsilon_{a b} Q_c Q_{\dot d} ,
\label{3b-9a}\\
	\left[  Q_a, Q_b Q_c, Q_{\dot d} \right]  &=&
  \mu_{1,5} \left( Q_a Q_b, x_{c \dot d} \right) +
  \mu_{2,5} \left( Q_a Q_c, x_{b \dot d} \right) +
  \mu_{3,5} \epsilon_{b c} Q_a Q_{\dot d} +
\nonumber \\
  &&
  \mu_{4,5} \left( Q_b Q_c, x_{a \dot d} \right) +
  \mu_{5,5} \epsilon_{a c} Q_b Q_{\dot d} +
  \mu_{6,5} \epsilon_{a b} Q_c Q_{\dot d} ,
\label{3b-9b}\\
	\left[ Q_a, Q_b, Q_c Q_{\dot d} \right]  &=&
  \nu_{1,5} \left( Q_a Q_b, x_{c \dot d} \right) +
  \nu_{2,5} \left( Q_a Q_c, x_{b \dot d} \right) +
  \nu_{3,5} \epsilon_{b c} Q_a Q_{\dot d} +
\nonumber \\
  &&
  \nu_{4,5} \left( Q_b Q_c, x_{a \dot d} \right) +
  \nu_{5,5} \epsilon_{a c} Q_b Q_{\dot d} +
  \nu_{6,5} \epsilon_{a b} Q_c Q_{\dot d} .
\label{3b-9c}
\end{eqnarray}

\subsubsection{Final form of the associators. Jacobiators}

Taking into account the relation \eqref{3-20} (as in Section \ref{without}), we obtain the following 4-point associators with one dot
\begin{eqnarray}
	\left[ Q_{\dot a} Q_b, Q_c, Q_d \right]  &=&
	- \frac{1}{2} \frac{\hbar}{\ell_0} \zeta_1
	\epsilon_{cd} Q_{\dot a} Q_b +
  \frac{1}{2} \frac{\hbar}{\ell_0}
  \tilde{\rho}_{6,2}
  \left( x_{b \dot a}	, Q_c Q_d \right)  ,
\label{3b-10}\\
	\left[  Q_{\dot a}, Q_b Q_c, Q_d \right]  &=& 0 ,
\label{3b-20}\\
	\left[  Q_{\dot a}, Q_b, Q_c Q_d \right]  &=&
	- \frac{1}{2} \frac{\hbar}{\ell_0} \zeta_1
	\epsilon_{cd} Q_{\dot a} Q_b -
	\frac{1}{2} \frac{\hbar}{\ell_0} \tilde{\rho}_{6,2}
\left( x_{b \dot a}	, Q_c Q_d \right) ,
\label{3b-30}\\
	\left[  Q_a Q_{\dot b}, Q_c, Q_d \right]  &=&
	\phantom{-} \frac{1}{2} \frac{\hbar}{\ell_0} \zeta_1
	\epsilon_{cd} 	Q_a Q_{\dot b} +
		\frac{1}{2} \frac{\hbar}{\ell_0} \tilde{\rho}_{6,3}
\left( x_{a \dot b}, 	Q_c Q_d \right) ,
\label{3b-40}\\
	\left[  Q_a, Q_{\dot b} Q_c, Q_d \right]  &=& 0 ,
\label{3b-50}\\
	\left[  Q_a, Q_{\dot b}, Q_c Q_d \right]  &=&
	\phantom{-} \frac{1}{2} \frac{\hbar}{\ell_0} \zeta_1
	\epsilon_{cd} 	Q_a Q_{\dot b} -
		\frac{1}{2} \frac{\hbar}{\ell_0} \tilde{\rho}_{6,3}
  \left( x_{a \dot b}, 	Q_c Q_d \right) ,
\label{3b-60}\\
	\left[ Q_a Q_b, Q_{\dot c}, Q_d \right]  &=&
  \phantom{-} \frac{1}{2} \frac{\hbar}{\ell_0}
  \zeta_1 \epsilon_{ab} Q_{\dot c} Q_d +
	\frac{1}{2} \frac{\hbar}{\ell_0} \tilde \rho_{1,4}
	\left( x_{d \dot c}, Q_a Q_b \right)  ,
\label{3b-70}\\
	\left[  Q_a, Q_b Q_{\dot c}, Q_d \right]  &=& 0 ,
\label{3b-80}\\
	\left[ Q_a, Q_b, Q_{\dot c} Q_d \right]  &=&
  \phantom{-} \frac{1}{2} \frac{\hbar}{\ell_0}
  \zeta_1 \epsilon_{ab} Q_{\dot c} Q_d -
	\frac{1}{2} \frac{\hbar}{\ell_0} \tilde \rho_{1,4}
	\left( x_{d \dot c}, Q_a Q_b \right) ,
\label{3b-90}\\
	\left[ Q_a Q_b, Q_c, Q_{\dot d} \right]  &=&
  - \frac{1}{2} \frac{\hbar}{\ell_0}
  \zeta_1 \epsilon_{ab} Q_c Q_{\dot d} +
	\frac{1}{2} \frac{\hbar}{\ell_0} \tilde \rho_{4,5}
	\left( x_{c \dot d}, Q_a Q_b \right) ,
\label{3b-100}\\
	\left[  Q_a, Q_b Q_c, Q_{\dot d} \right]  &=& 0 ,
\label{3b-110}\\
	\left[ Q_a, Q_b, Q_c Q_{\dot d} \right]  &=&
  - \frac{1}{2} \frac{\hbar}{\ell_0}
  \zeta_1 \epsilon_{ab} Q_c Q_{\dot d} -
  \frac{1}{2} \frac{\hbar}{\ell_0} \tilde \rho_{4,5}
	\left( x_{c \dot d}, Q_a Q_b \right) .
\label{3b-120}
\end{eqnarray}
Considering the relation \eqref{3-20}, we have found that
\begin{equation}
\zeta_2 = - \zeta_1  ,
\label{3b-130}
\end{equation}
and we set
\begin{equation}
	\left|  \tilde{\rho}_{6,2} \right|  = \left|  \tilde{\rho}_{6,3} \right| =
  \left|  \tilde{\rho}_{1,4} \right| = \left|  \tilde{\rho}_{4,5} \right| = 1.
\label{3b-140}
\end{equation}
One can check that the Jacobiators are
\begin{eqnarray}
	J\left( Q_{\dot a} Q_b, Q_c, Q_d \right) = 0
\label{3b-150}\\
	J\left( Q_a Q_{\dot b}, Q_c, Q_d \right) = 0
\label{3b-160}\\
	J\left( Q_a Q_b, Q_{\dot c}, Q_d \right) = 0
\label{3b-170}\\
	J\left( Q_a Q_b, Q_c, Q_{\dot d} \right) = 0
\label{3b-180}
\end{eqnarray}	
if
\begin{eqnarray}
	\tilde \rho_{1,4} &=& \tilde \rho_{6,2},
\label{3b-200}\\
	\tilde \rho_{4,5} &=& \tilde \rho_{6,3}.
\label{3b-210}
\end{eqnarray}	

\subsection{4-point associators with two dots}

In this section we consider the case with two dots and with the different locations.

\subsubsection{1-st case}

We seek 4-point associators with two dots as follows:
\begin{eqnarray}
	\left[ Q_{\dot a} Q_{\dot b}, Q_c, Q_d \right]  &=&
	\rho_{1,6} Q_{\dot a} Q_{\dot b} \epsilon_{c d} +
  \rho_{2,6} \left( Q_{\dot a} Q_{c}, y_{d \dot b} \right) +
  \rho_{3,6} \left( Q_{\dot a} Q_{d}, y_{c \dot b} \right) +
\nonumber \\
  &&
  \rho_{4,6} \left( Q_{\dot b} Q_{c}, y_{d \dot a} \right) +
  \rho_{5,6} \left( Q_{\dot b} Q_{d}, y_{c \dot a} \right) +
  \rho_{6,6} Q_{c} Q_{d} \epsilon_{\dot a \dot b} +
  \rho_{7,6} M_{\left( c \dot a, d \dot b \right)}  +
  \rho_{8,6} M_{\left( c \dot b, d \dot a \right)} ,
\label{3c1-10}\\
	\left[ Q_{\dot a}, Q_{\dot b} Q_c, Q_d \right]  &=&
	\mu_{1,6} Q_{\dot a} Q_{\dot b} \epsilon_{c d} +
  \mu_{2,6} \left( Q_{\dot a} Q_{c}, y_{d \dot b} \right) +
  \mu_{3,6} \left( Q_{\dot a} Q_{d}, y_{c \dot b} \right) +
\nonumber \\
  &&
  \mu_{4,6} \left( Q_{\dot b} Q_{c}, y_{d \dot a} \right) +
  \mu_{5,6} \left( Q_{\dot b} Q_{d}, y_{c \dot a} \right) +
  \mu_{6,6} Q_{c} Q_{d} \epsilon_{\dot a \dot b} +
  \mu_{7,6} M_{\left( c \dot a, d \dot b \right)}+
  \mu_{8,6} M_{\left( c \dot b, d \dot a \right)}  ,
\label{3c1-20}\\
	\left[ Q_{\dot a}, Q_{\dot b}, Q_c Q_d \right]  &=&
	\nu_{1,6} Q_{\dot a} Q_{\dot b} \epsilon_{c d} +
  \nu_{2,6} \left( Q_{\dot a} Q_{c}, y_{d \dot b} \right) +
  \nu_{3,6} \left( Q_{\dot a} Q_{d}, y_{c \dot b} \right) +
\nonumber \\
  &&
  \nu_{4,6} \left( Q_{\dot b} Q_{c}, y_{d \dot a} \right) +
  \nu_{5,6} \left( Q_{\dot b} Q_{d}, y_{c \dot a} \right) +
  \nu_{6,6} Q_{c} Q_{d} \epsilon_{\dot a \dot b} +
  \nu_{7,6} M_{\left( c \dot a, d \dot b \right)}  +
  \nu_{8,6} M_{\left( c \dot b, d \dot a \right)} .
\label{3c1-30}\\
\end{eqnarray}

\subsubsection{2-nd case}

We seek 4-point associators with two dots as follows:
\begin{eqnarray}
	\left[ Q_{\dot a} Q_b, Q_{\dot c}, Q_d \right]  &=&
	\rho_{1,7} \left( Q_{\dot a} Q_b, x_{d \dot c} \right) +
  \rho_{2,7} Q_{\dot a} Q_{\dot c} \epsilon_{b d} +
  \rho_{3,7} \left( Q_{\dot a} Q_d, x_{b \dot c} \right) +
\nonumber \\
  &&
  \rho_{4,7} \left( Q_b Q_{\dot c}, x_{d \dot a} \right) +
  \rho_{5,7} Q_b Q_d \epsilon_{\dot a \dot c} +
  \rho_{6,7} \left( Q_{\dot c} Q_d, x_{b \dot a} \right) +
  \rho_{7,7} M_{\left( b \dot a, d \dot c \right) } +
  \rho_{8,7} M_{\left( b \dot c, d \dot a \right) } ,
\label{3c2-10}\\
	\left[ Q_{\dot a}, Q_b Q_{\dot c}, Q_d \right]  &=&
	\mu_{1,7} \left( Q_{\dot a} Q_b, x_{d \dot c} \right) +
  \mu_{2,7} Q_{\dot a} Q_{\dot c} \epsilon_{b d} +
  \mu_{3,7} \left( Q_{\dot a} Q_d, x_{b \dot c} \right) +
\nonumber \\
  &&
  \mu_{4,7} \left( Q_b Q_{\dot c}, x_{d \dot a} \right) +
  \mu_{5,7} Q_b Q_d \epsilon_{\dot a \dot c} +
  \mu_{6,7} \left( Q_{\dot c} Q_d, x_{b \dot a} \right) +
  \mu_{7,7} M_{\left( b \dot a, d \dot c \right) } +
  \mu_{8,7} M_{\left( b \dot c, d \dot a \right) } ,
\label{3c2-20}\\
	\left[ Q_{\dot a}, Q_b, Q_{\dot c} Q_d \right]  &=&
	\nu_{1,7} \left( Q_{\dot a} Q_b, x_{d \dot c} \right) +
  \nu_{2,7} Q_{\dot a} Q_{\dot c} \epsilon_{b d} +
  \nu_{3,7} \left( Q_{\dot a} Q_d, x_{b \dot c} \right) +
\nonumber \\
  &&
  \nu_{4,7} \left( Q_b Q_{\dot c}, x_{d \dot a} \right) +
  \nu_{5,7} Q_b Q_d \epsilon_{\dot a \dot c} +
  \nu_{6,7} \left( Q_{\dot c} Q_d, x_{b \dot a} \right) +
  \nu_{7,7} M_{\left( b \dot a, d \dot c \right) } +
  \nu_{8,7} M_{\left( b \dot c, d \dot a \right) }.
\label{3c2-30}
\end{eqnarray}

\subsubsection{ 3-rd case}

We seek 4-point associators with two dots as follows:
\begin{eqnarray}
	\left[ Q_{\dot a} Q_b, Q_{c}, Q_{\dot d} \right]  &=&
	\rho_{1,8} \left( Q_{\dot a} Q_b, x_{c \dot d} \right) +
  \rho_{2,8} \left( Q_{\dot a} Q_c, x_{b \dot d} \right) +
  \rho_{3,8} Q_{\dot a} Q_{\dot d} \epsilon_{b c} +
\nonumber \\
  &&
  \rho_{4,8} \left( Q_b Q_c \epsilon_{\dot a \dot d} \right) +
  \rho_{5,8} \left( Q_b Q_{\dot d}, x_{c \dot a} \right) +
  \rho_{6,8} \left( Q_c Q_{\dot d}, x_{b \dot a} \right) +
  \rho_{7,8} M_{\left( b \dot a, c \dot d \right) } +
  \rho_{8,8} M_{\left( b \dot d, c \dot a \right) } ,
\label{3c3-10}\\
	\left[ Q_{\dot a}, Q_b Q_{c}, Q_{\dot d} \right]  &=&
	\mu_{1,8} \left( Q_{\dot a} Q_b, x_{c \dot d} \right) +
  \mu_{2,8} \left( Q_{\dot a} Q_c, x_{b \dot d} \right) +
  \mu_{3,8} Q_{\dot a} Q_{\dot d} \epsilon_{b c} +
\nonumber \\
  &&
  \mu_{4,8} \left( Q_b Q_c \epsilon_{\dot a \dot d} \right) +
  \mu_{5,8} \left( Q_b Q_{\dot d}, x_{c \dot a} \right) +
  \mu_{6,8} \left( Q_c Q_{\dot d}, x_{b \dot a} \right) +
  \mu_{7,8} M_{\left( b \dot a, c \dot d \right) } +
  \mu_{8,8} M_{\left( b \dot d, c \dot a \right) } ,
\label{3c3-20}\\
	\left[ Q_{\dot a}, Q_b, Q_{c} Q_{\dot d} \right]  &=&
	\nu_{1,8} \left( Q_{\dot a} Q_b, x_{c \dot d} \right) +
  \nu_{2,8} \left( Q_{\dot a} Q_c, x_{b \dot d} \right) +
  \nu_{3,8} Q_{\dot a} Q_{\dot d} \epsilon_{b c} +
\nonumber \\
  &&
  \nu_{4,8} \left( Q_b Q_c \epsilon_{\dot a \dot d} \right) +
  \nu_{5,8} \left( Q_b Q_{\dot d}, x_{c \dot a} \right) +
  \nu_{6,8} \left( Q_c Q_{\dot d}, x_{b \dot a} \right) +
  \nu_{7,8} M_{\left( b \dot a, c \dot d \right) } +
  \nu_{8,8} M_{\left( b \dot d, c \dot a \right) }.
\label{3c3-30}
\end{eqnarray}

\subsubsection{4-th case}

We seek 4-point associators with two dots as follows:
\begin{eqnarray}
	\left[ Q_{a} Q_{\dot b}, Q_{\dot c}, Q_{d} \right]  &=&
	\rho_{1,9} \left( Q_{a} Q_{\dot b}, x_{d \dot c} \right) +
  \rho_{2,9} \left( Q_{a} Q_{\dot c}, x_{d \dot b} \right) +
  \rho_{3,9} \epsilon_{\dot b \dot c} Q_{a} Q_{d} +
\nonumber \\
  &&
  \rho_{4,9} \epsilon_{a d} Q_{\dot b} Q_{\dot c} +
  \rho_{5,9} \left( Q_{\dot b} Q_{d}, x_{a \dot c} \right) +
  \rho_{6,9} \left( Q_{\dot c} Q_{d}, x_{a \dot b} \right) +
  \rho_{7,9} M_{\left( a \dot b, d \dot c \right) } +
  \rho_{8,9} M_{\left( a \dot c, d \dot b \right) } ,
\label{3c4-10}\\
	\left[ Q_{a}, Q_{\dot b} Q_{\dot c}, Q_{d} \right]  &=&
	\mu_{1,9} \left( Q_{a} Q_{\dot b}, x_{d \dot c} \right) +
  \mu_{2,9} \left( Q_{a} Q_{\dot c}, x_{d \dot b} \right) +
  \mu_{3,9} \epsilon_{\dot b \dot c} Q_{a} Q_{d} +
\nonumber \\
  &&
 \mu_{4,9} \epsilon_{a d} Q_{\dot b} Q_{\dot c} +
  \mu_{5,9} \left( Q_{\dot b} Q_{d}, x_{a \dot c} \right) +
  \mu_{6,9} \left( Q_{\dot c} Q_{d}, x_{a \dot b} \right) +
  \mu_{7,9} M_{\left( a \dot b, d \dot c \right) } +
  \mu_{8,9} M_{\left( a \dot c, d \dot b \right) } ,
\label{3c4-20}\\
	\left[ Q_{a}, Q_{\dot b}, Q_{\dot c} Q_{d} \right]  &=&
	\nu_{1,9} \left( Q_{a} Q_{\dot b}, x_{d \dot c} \right) +
  \nu_{2,9} \left( Q_{a} Q_{\dot c}, x_{d \dot b} \right) +
  \nu_{3,9} \epsilon_{\dot b \dot c} Q_{a} Q_{d} +
\nonumber \\
  &&
  \nu_{4,9} \epsilon_{a d} Q_{\dot b} Q_{\dot c} +
  \nu_{5,9} \left( Q_{\dot b} Q_{d}, x_{a \dot c} \right) +
  \nu_{6,9} \left( Q_{\dot c} Q_{d}, x_{a \dot b} \right) +
  \nu_{7,9} M_{\left( a \dot b, d \dot c \right) } +
  \nu_{8,9} M_{\left( a \dot c, d \dot b \right) } .
\label{3c4-30}
\end{eqnarray}

\subsubsection{5-th case}

We seek 4-point associators with two dots as follows:
\begin{eqnarray}
	\left[ Q_{a} Q_{\dot b}, Q_{c}, Q_{\dot d} \right]  &=&
	\rho_{1,10} \left( Q_{a} Q_{\dot b}, x_{c \dot d} \right) +
  \rho_{2,10} \epsilon_{\dot b \dot d} Q_{a} Q_{c} +
  \rho_{3,10} \left( Q_{a} Q_{\dot d}, x_{c \dot b} \right) +
\nonumber \\
  &&
  \rho_{4,10} \left( Q_{\dot b} Q_{c}, x_{a \dot d} \right) +
  \rho_{5,10} \epsilon_{a c} Q_{\dot b} Q_{\dot d} +
  \rho_{6,10} \left( Q_{c} Q_{\dot d}, x_{a \dot b} \right) +
  \rho_{7,10} M_{\left( a \dot b, c \dot d \right) } +
  \rho_{8 7} M_{\left( a \dot d, c \dot b \right) } ,
\label{3c5-10}\\
	\left[ Q_{a}, Q_{\dot b} Q_{c}, Q_{\dot d} \right]  &=&
	\mu_{1,10} \left( Q_{a} Q_{\dot b}, x_{c \dot d} \right) +
  \mu_{2,10} \epsilon_{\dot b \dot d} Q_{a} Q_{c} +
  \mu_{3,10} \left( Q_{a} Q_{\dot d}, x_{c \dot b} \right) +
\nonumber \\
  &&
  \mu_{4,10} \left( Q_{\dot b} Q_{c}, x_{a \dot d} \right) +
  \mu_{5,10} \epsilon_{a c} Q_{\dot b} Q_{\dot d} +
  \mu_{6,10} \left( Q_{c} Q_{\dot d}, x_{a \dot b} \right) +
  \mu_{7,10} M_{\left( a \dot b, c \dot d \right) } +
  \mu_{8,10} M_{\left( a \dot d, c \dot b \right) } ,
\label{3c5-20}\\
	\left[ Q_{a}, Q_{\dot b}, Q_{c} Q_{\dot d} \right]  &=&
	\nu_{1,10} \left( Q_{a} Q_{\dot b}, x_{c \dot d} \right) +
  \nu_{2,10} \epsilon_{\dot b \dot d} Q_{a} Q_{c} +
  \nu_{3,10} \left( Q_{a} Q_{\dot d}, x_{c \dot b} \right) +
\nonumber \\
  &&
  \nu_{4,10} \left( Q_{\dot b} Q_{c}, x_{a \dot d} \right) +
  \nu_{5,10} \epsilon_{a c} Q_{\dot b} Q_{\dot d} +
  \nu_{6,10} \left( Q_{c} Q_{\dot d}, x_{a \dot b} \right) +
  \nu_{7,10} M_{\left( a \dot b, c \dot d \right) } +
  \nu_{8,10} M_{\left( a \dot d, c \dot b \right) } .
\label{3c5-30}
\end{eqnarray}

\subsubsection{6-th case}

We seek 4-point associators with two dots as follows:
\begin{eqnarray}
	\left[ Q_{a} Q_{b}, Q_{\dot c}, Q_{\dot d} \right]  &=&
	\rho_{1,11} \epsilon_{\dot c \dot d} Q_{a} Q_{b}  +
  \rho_{2,11} \left( Q_{a} Q_{\dot c}, x_{b \dot d} \right)+
  \rho_{3,11} \left( Q_{a} Q_{\dot d}, x_{b \dot c} \right) +
\nonumber \\
  &&
  \rho_{4,11} \left( Q_{b} Q_{\dot c}, x_{a \dot d} \right) +
  \rho_{5,11} \left( Q_{b} Q_{\dot d}, x_{a \dot c} \right) +
  \rho_{6,11} \epsilon_{a b} Q_{\dot c} Q_{\dot d} +
  \rho_{7,11} M_{\left( a \dot c, b \dot d \right) } +
  \rho_{8,11} M_{\left( a \dot d, b \dot c \right) } ,
\label{3c6-10}\\
	\left[ Q_{a}, Q_{b} Q_{\dot c}, Q_{\dot d} \right]  &=&
	\mu_{1,11} \epsilon_{\dot c \dot d} Q_{a} Q_{b}  +
  \mu_{2,11} \left( Q_{a} Q_{\dot c}, x_{b \dot d} \right)+
  \mu_{3,11} \left( Q_{a} Q_{\dot d}, x_{b \dot c} \right) +
\nonumber \\
  &&
  \mu_{4,11} \left( Q_{b} Q_{\dot c}, x_{a \dot d} \right) +
  \mu_{5,11} \left( Q_{b} Q_{\dot d}, x_{a \dot c} \right) +
  \mu_{6,11} \epsilon_{a b} Q_{\dot c} Q_{\dot d} +
  \mu_{7,11} M_{\left( a \dot c, b \dot d \right) } +
  \mu_{8,11} M_{\left( a \dot d, b \dot c \right) } ,
\label{3c6-20}\\
	\left[ Q_{a}, Q_{b}, Q_{\dot c} Q_{\dot d} \right]  &=&
	\nu_{1,11} \epsilon_{\dot c \dot d} Q_{a} Q_{b}  +
  \nu_{2,11} \left( Q_{a} Q_{\dot c}, x_{b \dot d} \right)+
  \nu_{3,11} \left( Q_{a} Q_{\dot d}, x_{b \dot c} \right) +
\nonumber \\
  &&
  \nu_{4,11} \left( Q_{b} Q_{\dot c}, x_{a \dot d} \right) +
  \nu_{5,11} \left( Q_{b} Q_{\dot d}, x_{a \dot c} \right) +
  \nu_{6,11} \epsilon_{a b} Q_{\dot c} Q_{\dot d} +
  \nu_{7,11} M_{\left( a \dot c, b \dot d \right) } +
  \nu_{8,11} M_{\left( a \dot d, b \dot c \right) }  .
\label{3c6-30}
\end{eqnarray}

\subsubsection{Final form of the associators. Jacobiators}

Using the relation \eqref{3-20} for the connection between 3- and 4-point associators and the simplifications similar to those of used in section \ref{without}, we obtain
\begin{eqnarray}
	\left[ Q_{\dot a} Q_{\dot b}, Q_c, Q_d \right]  &=&
  - \frac{\hbar}{\ell_0} \zeta_2 Q_{\dot a} Q_{\dot b} \epsilon_{c d} +
  \frac{\hbar}{\ell_0} \tilde \rho_{7,6} M_{\left( c \dot a, d \dot b \right)} +
  \frac{\hbar}{\ell_0} \tilde \rho_{8,6} M_{\left( c \dot b, d \dot a \right)} ,
\label{3c1-40}\\
	\left[ Q_{\dot a}, Q_{\dot b} Q_c, Q_d \right]  &=&
	\frac{\hbar}{\ell_0} \tilde \mu_{7,6} M_{\left( c \dot a, d \dot b \right)} +
  \frac{\hbar}{\ell_0} \tilde \mu_{8,6} M_{\left( c \dot b, d \dot a \right)}  ,
\label{3c1-50}\\
	\left[ Q_{\dot a}, Q_{\dot b}, Q_c Q_d \right]  &=&
	- \frac{\hbar}{\ell_0} \zeta_3 Q_{c} Q_{d} \epsilon_{\dot a \dot b} + \frac{\hbar}{\ell_0} \tilde \nu_{7,6} M_{\left( c \dot a, d \dot b \right)} +
  \frac{\hbar}{\ell_0} \tilde \nu_{8,6} M_{\left( c \dot b, d \dot a \right)},
\label{3c1-60}\\
	\left[ Q_{\dot a} Q_b, Q_{\dot c}, Q_d \right]  &=&
  \frac{1}{2} \frac{\hbar}{\ell_0} \tilde \rho_{1,7}
  \left( Q_{\dot a} Q_b, x_{d \dot c} \right) +
  \frac{1}{2} \frac{\hbar}{\ell_0} \tilde \rho_{6,7}
  \left( x_{b \dot a}, Q_{\dot c} Q_d \right) +
  \frac{\hbar}{\ell_0} \tilde \rho_{7,7} M_{\left( b \dot a, d \dot c \right) } +
  \frac{\hbar}{\ell_0} \tilde \rho_{8,7} M_{\left( b \dot c, d \dot a \right) } ,
\label{3c2-40}\\
	\left[ Q_{\dot a}, Q_b Q_{\dot c}, Q_d \right]  &=&
	\frac{\hbar}{\ell_0} \tilde \mu_{7,7} M_{\left( b \dot a, d \dot c \right) } +
  \frac{\hbar}{\ell_0} \tilde \mu_{8,7} M_{\left( b \dot c, d \dot a \right) } ,
\label{3c2-50}\\
	\left[ Q_{\dot a}, Q_b, Q_{\dot c} Q_d \right]  &=&
	- \frac{1}{2} \frac{\hbar}{\ell_0} \tilde \rho_{1,7}
  \left( Q_{\dot a} Q_b, x_{d \dot c} \right) -
  \frac{1}{2} \frac{\hbar}{\ell_0} \tilde \rho_{6,7}
  \left( x_{b \dot a}, Q_{\dot c} Q_d \right) +
  \frac{\hbar}{\ell_0} \tilde \nu_{7,7} M_{\left( b \dot a, d \dot c \right) } +
  \frac{\hbar}{\ell_0} \tilde \nu_{8,7} M_{\left( b \dot c, d \dot a \right) } ,
\label{3c2-60}\\
	\left[ Q_{\dot a} Q_b, Q_{c}, Q_{\dot d} \right]  &=&
  \frac{\hbar}{\ell_0} \tilde \rho_{1,8} \left( Q_{\dot a} Q_b, x_{c \dot d} \right) +
  \frac{\hbar}{\ell_0} \zeta_2 Q_{\dot a} Q_{\dot d} \epsilon_{b c} +
  \frac{\hbar}{\ell_0} \tilde \rho_{4,8} \epsilon_{\dot a \dot d} Q_b Q_c +
  \frac{\hbar}{\ell_0} \tilde \rho_{6,8} \left( Q_c Q_{\dot d}, x_{b \dot a} \right) +
\nonumber \\
  &&
  \frac{\hbar}{\ell_0} \tilde \rho_{7,8} M_{\left( b \dot a, c \dot d \right) } +
  \frac{\hbar}{\ell_0} \tilde \rho_{8,8} M_{\left( b \dot d, c \dot a \right) } ,
\label{3c3-40}\\
	\left[ Q_{\dot a}, Q_b Q_{c}, Q_{\dot d} \right]  &=&
-2 \frac{\hbar}{\ell_0} \tilde \rho_{4,8} \epsilon_{\dot a \dot d} Q_b Q_c +
	\tilde \mu_{7,8} M_{\left( b \dot a, c \dot d \right) } +
  \tilde \mu_{8,8} M_{\left( b \dot d, c \dot a \right) } ,
\label{3c3-50}\\
	\left[ Q_{\dot a}, Q_b, Q_{c} Q_{\dot d} \right]  &=&
	- \frac{\hbar}{\ell_0} \tilde \rho_{1,8} \left( Q_{\dot a} Q_b, x_{c \dot d} \right) +
  \frac{\hbar}{\ell_0} \zeta_2 Q_{\dot a} Q_{\dot d} \epsilon_{b c} +
  \frac{\hbar}{\ell_0} \tilde \rho_{4,8} \epsilon_{\dot a \dot d} Q_b Q_c -
  \frac{\hbar}{\ell_0} \tilde \rho_{6,8} \left( Q_c Q_{\dot d}, x_{b \dot a} \right) +
\nonumber \\
  &&
  \frac{\hbar}{\ell_0} \tilde \nu_{7,8} M_{\left( b \dot a, c \dot d \right) } +
  \frac{\hbar}{\ell_0} \tilde \nu_{8,8} M_{\left( b \dot d, c \dot a \right) } ,
\label{3c3-60}\\
	\left[ Q_{a} Q_{\dot b}, Q_{\dot c}, Q_d \right]  &=&
  \frac{\hbar}{\ell_0} \tilde \rho_{1,9} \left( Q_{a} Q_{\dot b}, x_{d \dot c} \right) +
  \frac{\hbar}{\ell_0} \zeta_3 \epsilon_{\dot b \dot c} Q_{a} Q_{d} +
  \frac{\hbar}{\ell_0} \tilde \rho_{4,9} \epsilon_{ad} Q_{\dot b} Q_{\dot c} +
  \frac{\hbar}{\ell_0} \tilde \rho_{6,9} \left( Q_{\dot c} Q_{d}, x_{a \dot b} \right) +
\nonumber \\
  &&
  \frac{\hbar}{\ell_0} \tilde \rho_{7,9} M_{\left( a \dot b, d \dot c \right) } +
  \frac{\hbar}{\ell_0} \tilde \rho_{8,9} M_{\left( a \dot c, d \dot b \right) } ,
\label{3c4-40}\\
	\left[ Q_{a}, Q_{\dot b} Q_{\dot c}, Q_d \right]  &=&
	 - 2 \frac{\hbar}{\ell_0} \tilde \rho_{4,9} \epsilon_{ad} Q_{\dot b} Q_{\dot c} +
  \frac{\hbar}{\ell_0} \tilde \mu_{7,9} M_{\left( a \dot b, d \dot c \right) } +
  \frac{\hbar}{\ell_0} \tilde \mu_{8,9} M_{\left( a \dot c, d \dot b \right) } ,
\label{3c4-50}\\
	\left[ Q_{a}, Q_{\dot b}, Q_{\dot c} Q_d \right]  &=&
	- \frac{\hbar}{\ell_0} \tilde \rho_{1,9} \left( Q_{a} Q_{\dot b}, x_{d \dot c} \right) +
  \frac{\hbar}{\ell_0} \zeta_3 \epsilon_{\dot b \dot c} Q_{a} Q_{d} +
  \frac{\hbar}{\ell_0} \tilde \rho_{4,9} \epsilon_{ad} Q_{\dot b} Q_{\dot c} -
  \frac{\hbar}{\ell_0} \tilde \rho_{6,9} \left( Q_{\dot c} Q_{d}, x_{a \dot b} \right) +
\nonumber \\
  &&
  \frac{\hbar}{\ell_0} \tilde \nu_{7,9} M_{\left( a \dot b, d \dot c \right) } +
  \frac{\hbar}{\ell_0} \tilde \nu_{8,9} M_{\left( a \dot c, d \dot b \right) } ,
\label{3c4-60}\\
	\left[ Q_{a} Q_{\dot b}, Q_{c}, Q_{\dot d} \right]  &=&
  \frac{\hbar}{\ell_0} \tilde \rho_{1,10} \left( Q_{a} Q_{\dot b}, x_{c \dot d} \right) +
  \frac{\hbar}{\ell_0} \tilde \rho_{6,10} \left( x_{a \dot b}, Q_{c} Q_{\dot d} \right) +
  \frac{\hbar}{\ell_0} \tilde \rho_{7,10} M_{\left( a \dot b, d \dot c \right) } +
  \frac{\hbar}{\ell_0} \tilde \rho_{8,10} M_{\left( a \dot c, d \dot b \right) } ,
\label{3c5-40}\\
	\left[ Q_{a}, Q_{\dot b} Q_{c}, Q_{\dot d} \right]  &=&
	\frac{\hbar}{\ell_0} \tilde \mu_{7,10} M_{\left( a \dot b, c \dot d \right) } +
  \frac{\hbar}{\ell_0} \tilde \mu_{8,10} M_{\left( a \dot d, c \dot b \right) }  ,
\label{3c5-50}\\
	\left[ Q_{a}, Q_{\dot b}, Q_{c} Q_{\dot d} \right]  &=&
	- \frac{\hbar}{\ell_0} \tilde \rho_{1,10} \left( Q_{a} Q_{\dot b}, x_{c \dot d} \right) -
  \frac{\hbar}{\ell_0} \tilde \rho_{6,10} \left( x_{a \dot b}, Q_{c} Q_{\dot d} \right) +
  \frac{\hbar}{\ell_0} \tilde \nu_{7,10} M_{\left( a \dot b, d \dot c \right) } +
  \frac{\hbar}{\ell_0} \tilde \nu_{8,10} M_{\left( a \dot c, d \dot b \right) } ,
\label{3c5-60}\\
	\left[ Q_{a} Q_{b}, Q_{\dot c}, Q_{\dot d} \right]  &=&
  - \frac{\hbar}{\ell_0} \zeta_3 \epsilon_{\dot c \dot d} Q_{a} Q_{b} +
  \frac{\hbar}{\ell_0} \tilde \rho_{7,11} M_{\left( a \dot c, b \dot d \right) } +
  \frac{\hbar}{\ell_0} \tilde \rho_{8,11} M_{\left( a \dot d, b \dot c \right) } ,
\label{3c6-40}\\
	\left[ Q_{a}, Q_{b} Q_{\dot c}, Q_{\dot d} \right]  &=&
	\frac{\hbar}{\ell_0} \tilde \mu_{7,11} M_{\left( a \dot c, b \dot d \right) } +
  \frac{\hbar}{\ell_0} \tilde \mu_{8,11} M_{\left( a \dot d, b \dot c \right) } ,
\label{3c6-50}\\
	\left[ Q_{a}, Q_{b}, Q_{\dot c} Q_{\dot d} \right]  &=&
	- \frac{\hbar}{\ell_0} \zeta_2 \epsilon_{a b} Q_{\dot c} Q_{\dot d} +
  \frac{\hbar}{\ell_0} \tilde \nu_{7,11} M_{\left( a \dot c, b \dot d \right) } +
  \frac{\hbar}{\ell_0} \tilde \nu_{8,11} M_{\left( a \dot d, b \dot c \right) } ,
\label{3c6-60}
\end{eqnarray}
where $(\ldots , \ldots)$ is either commutator or anticommutator, and with the limitations
\begin{eqnarray}
  \tilde \rho_{7,6} -  \tilde \mu_{7,6} +  \tilde \nu_{7,6} = 0 ,
\label{3c1-70}\\
  \tilde \rho_{8,6} -  \tilde \mu_{8,6} +  \tilde \nu_{8,6} = 0 ,
\label{3c1-80}\\
  \tilde \rho_{7,7} -  \tilde \mu_{7,7} +  \tilde \nu_{7,7} = 0 ,
\label{3c2-90}\\
  \tilde \rho_{8,7} -  \tilde \mu_{8,7} +  \tilde \nu_{8,7} = 0 ,
\label{3c2-100}\\
  \tilde \rho_{7,8} -  \tilde \mu_{7,8} +  \tilde \nu_{7,8} = 0 ,
\label{3c3-110}\\
  \tilde \rho_{8,8} -  \tilde \mu_{8,8} +  \tilde \nu_{8,8} = 0 ,
\label{3c3-120}\\
  \tilde \rho_{7,9} -  \tilde \mu_{7,9} +  \tilde \nu_{7,9} = 0 ,
\label{3c4-130}\\
  \tilde \rho_{8,9} -  \tilde \mu_{8,9} +  \tilde \nu_{8,9} = 0 ,
\label{3c4-150}\\
  \tilde \rho_{7,10} -  \tilde \mu_{7,10} +  \tilde \nu_{7,10} = 0 ,
\label{3c5-160}\\
  \tilde \rho_{8,10} -  \tilde \mu_{8,10} +  \tilde \nu_{8,10} = 0 ,
\label{3c5-170}\\
  \tilde \rho_{7,11} -  \tilde \mu_{7,11} +  \tilde \nu_{7,11} = 0 ,
\label{3c6-180}\\
  \tilde \rho_{8,11} -  \tilde \mu_{8,11} +  \tilde \nu_{8,11} = 0 .
\label{3c6-190}
\end{eqnarray}
One can check that the Jacobiators are (here we  consider the simplest case
$x_{a, \dot b} = M_{\left( a \dot b, c \dot d \right)} = 0$)
\begin{eqnarray}
	J\left( Q_{\dot a} Q_{\dot b}, Q_c, Q_d \right) &=& 0 ,
\label{3c-200}\\
	J\left( Q_{\dot a}, Q_{\dot b} Q_c, Q_d \right) &=&
  - \frac{\hbar}{\ell_0} \zeta_2 \epsilon_{cd} \left[ Q_{\dot a}, Q_{\dot b} \right] -
  \frac{\hbar}{\ell_0} \zeta_3 \epsilon_{\dot a \dot b} \left[ Q_{c}, Q_{d} \right] ,
\label{3c-210}\\
	J\left( Q_{\dot a}, Q_{\dot b}, Q_c Q_d \right) &=& 0 ,
 \label{3c-210a}\\
	J\left( Q_{\dot a} Q_b, Q_{\dot c}, Q_d \right) &=&
   - \frac{\hbar}{\ell_0} \zeta_2 \epsilon_{bd} \left[ Q_{\dot a}, Q_{\dot c} \right] -
  \frac{\hbar}{\ell_0} \zeta_3 \epsilon_{\dot a \dot c} \left[ Q_{b}, Q_{d} \right] ,
\label{3c-220}\\
  J\left( Q_{\dot a}, Q_b Q_{\dot c}, Q_d \right) &=&
  \frac{\hbar}{\ell_0} \zeta_2 \epsilon_{bd} \left[ Q_{\dot a}, Q_{\dot c} \right] +
  \frac{\hbar}{\ell_0} \zeta_3 \epsilon_{\dot a \dot c} \left[ Q_{b}, Q_{d} \right] ,
\label{3c-230}\\
  J\left( Q_{\dot a}, Q_b, Q_{\dot c} Q_d \right) &=&
  - \frac{\hbar}{\ell_0} \zeta_2 \epsilon_{bd} \left[ Q_{\dot a}, Q_{\dot c} \right] -
  \frac{\hbar}{\ell_0} \zeta_3 \epsilon_{\dot a \dot c} \left[ Q_{b}, Q_{d} \right] ,
\label{3c-240}\\
  J\left( Q_{\dot a} Q_b, Q_{c} Q_{\dot d} \right) &=&
  \frac{\hbar}{\ell_0} \zeta_2 \epsilon_{bc} \left[ Q_{\dot a}, Q_{\dot d} \right] +
  \frac{\hbar}{\ell_0} \zeta_3 \epsilon_{\dot a \dot d} \left[ Q_{b}, Q_{c} \right] ,
\label{3c-250}\\
  J\left( Q_{\dot a}, Q_b Q_{c}, Q_{\dot d} \right) &=& 0 ,
\label{3c-260}\\
  J\left( Q_{\dot a}, Q_b, Q_{c} Q_{\dot d} \right) &=&
  \frac{\hbar}{\ell_0} \zeta_2 \epsilon_{bc} \left[ Q_{\dot a}, Q_{\dot d} \right] +
  \frac{\hbar}{\ell_0} \zeta_3 \epsilon_{\dot a \dot d} \left[ Q_{b}, Q_{c} \right] ,
\label{3c-270}\\
  J\left( Q_{a} Q_{\dot b}, Q_{\dot c}, Q_{d} \right) &=&
  \frac{\hbar}{\ell_0} \zeta_2 \epsilon_{ad} \left[ Q_{\dot b}, Q_{\dot c} \right] +
  \frac{\hbar}{\ell_0} \zeta_3 \epsilon_{\dot b \dot c} \left[ Q_{a}, Q_{d} \right] ,
\label{3c-280}\\
  J\left( Q_{a}, Q_{\dot b} Q_{\dot c}, Q_{d} \right) &=& 0 ,
\label{3c-290}\\
  J\left( Q_{a}, Q_{\dot b}, Q_{\dot c} Q_{d} \right) &=&
  \frac{\hbar}{\ell_0} \zeta_2 \epsilon_{ad} \left[ Q_{\dot b}, Q_{\dot c} \right] +
  \frac{\hbar}{\ell_0} \zeta_3 \epsilon_{\dot b \dot c} \left[ Q_{a}, Q_{d} \right] ,
\label{3c-300}\\
  J\left( Q_{a} Q_{\dot b}, Q_{c}, Q_{\dot d} \right) &=&
  - \frac{\hbar}{\ell_0} \zeta_2 \epsilon_{ac} \left[ Q_{\dot b}, Q_{\dot d} \right] -
  \frac{\hbar}{\ell_0} \zeta_3 \epsilon_{\dot b \dot d} \left[ Q_{a}, Q_{c} \right] ,
\label{3c-310}\\
  J\left( Q_{a}, Q_{\dot b} Q_{c}, Q_{\dot d} \right) &=&
  \frac{\hbar}{\ell_0} \zeta_2 \epsilon_{ac} \left[ Q_{\dot b}, Q_{\dot d} \right] +
  \frac{\hbar}{\ell_0} \zeta_3 \epsilon_{\dot b \dot d} \left[ Q_{a}, Q_{c} \right] ,
\label{3c-320}\\
  J\left( Q_{a}, Q_{\dot b}, Q_{c} Q_{\dot d} \right) &=&
  - \frac{\hbar}{\ell_0} \zeta_2 \epsilon_{ac} \left[ Q_{\dot b}, Q_{\dot d} \right] -
  \frac{\hbar}{\ell_0} \zeta_3 \epsilon_{\dot b \dot d} \left[ Q_{a}, Q_{c} \right] ,
\label{3c-330}\\
  J\left( Q_{a} Q_{b}, Q_{\dot c} Q_{\dot d} \right) &=& 0 ,
\label{3c-340}\\
  J\left( Q_{a}, Q_{b} Q_{\dot c}, Q_{\dot d} \right) &=&
  - \frac{\hbar}{\ell_0} \zeta_2 \epsilon_{ab} \left[ Q_{\dot c}, Q_{\dot d} \right] -
  \frac{\hbar}{\ell_0} \zeta_3 \epsilon_{\dot c \dot d} \left[ Q_{a}, Q_{b} \right] ,
\label{3c-350}\\
  J\left( Q_{a}, Q_{b}, Q_{\dot c} Q_{\dot d} \right) &=& 0 ,
\label{3c-360}\\
\end{eqnarray}	
if
\begin{eqnarray}
	\tilde \rho_{4,8} &=& \zeta_3 ,
\label{3c-370}\\
	\tilde \rho_{4,9} &=& \zeta_2 ,
\label{3c-380}\\
	\left\{ Q_{\dot a}, Q_{\dot b} \right\} &=& \left\{ Q_{a}, Q_{b} \right\} = 0 .
\label{3c-390}
\end{eqnarray}	

\subsection{4-point associators with two dots}

This case is identical to subsection \ref{3b} after replacing
$\dot a \leftrightarrow a$.

\section{The connection with supersymmetry}

Now we want to pounce on  supersymmetry. In this case the operators
$Q_{a, \dot b}$ obey the following anticommutators
\begin{eqnarray}
	\left\{
	Q_a , Q_b
	\right\} &=& \left\{
	Q_{\dot a} , Q_{\dot b}
	\right\} = 0,
\label{4-10} \\
  \left\{
  Q_a , Q_{\dot a}
  \right\} &=& Q_a Q_{\dot a} + Q_{\dot a} Q_a =
  2 \sigma^\mu_{a \dot a} P_\mu
\label{4-20},
\end{eqnarray}
where the operator $P$ can be a nonassociative generalization of standard accosiative operator $- i \hbar \partial_\mu$;
the Pauli matrices $\sigma^\mu_{a \dot a}, \sigma_\mu^{a \dot a}$ are defined in the standard way
\begin{eqnarray}
	\sigma^\mu_{a \dot a} &=& \left\{
	\left(
	\begin{array}{cc}
	1 & 0 \\
	0 & 1 \\
	\end{array}
	\right),
	\left(
	\begin{array}{cc}
	0 & 1 \\
	1 & 0 \\
	\end{array}
	\right),
	\left(
	\begin{array}{cc}
	0 & -i \\
	i & 0 \\
	\end{array}
	\right),
	\left(
	\begin{array}{cc}
	1 & 0 \\
	0 & -1 \\
	\end{array}
	\right)
	\right\}
\label{4-30}\\
	\sigma_\mu^{a \dot a} &=& \left\{
	\left(
	\begin{array}{cc}
	1 & 0 \\
	0 & 1 \\
	\end{array}
	\right),
	\left(
	\begin{array}{cc}
	0 & 1 \\
	1 & 0 \\
	\end{array}
	\right),
	\left(
	\begin{array}{cc}
	0 & i \\
	-i & 0 \\
	\end{array}
	\right),
	\left(
	\begin{array}{cc}
	1 & 0 \\
	0 & -1 \\
	\end{array}
	\right)
	\right\}
\label{4-40}
\end{eqnarray}
with the orthogonality relations for the Pauli matrices
\begin{equation}\label{4-50}
	\sigma_\mu^{a \dot a} \sigma^\nu_{a \dot a} = 2 \delta_\mu^\nu, \quad
	\sigma_\mu^{a \dot a} \sigma^\mu_{b \dot b} =
	2 \delta_b^a \delta_{\dot b}^{\dot a}.
\end{equation}
Let us note the following interesting relation which follows from \eqref{2-70a}
\begin{equation}
	\left( Q_a Q_b \right) Q_b =
	- \frac{\hbar}{\ell_0} \zeta_1 Q_b \epsilon_{ab} .
\label{4-60}
\end{equation}
Roughly speaking, one can say that this expression ``destroys'' sometimes $Q_a^2=0$ property of Grassmann numbers.
This results in distinctions between supersymmetries based on associative and nonassociative generators.
But this difference will be (in the dimensionless form) of the order of $l_{Pl}/\ell_0$, where $\ell_0$ is some characteristic length. For example, if
$\ell_0 = \Lambda^{-1/2}$ (where $\Lambda$ is the cosmological constant) then this difference will be $\approx 10^{-120}$.

\section{Supersymmetry, hidden variables, and nonassociativity}

In this section we want to consider a possible connection between supersymmetry, hidden variables, and nonassociativity.

First of all we want to remind what is the hidden variables theory. In Wiki \cite{wikiHVT} one can find the following definition of hidden variables theories
''\dots hidden variable theories were espoused by some physicists who argued that the state of a physical system, as formulated by quantum mechanics,
does not give a complete description for the system; i.e., that quantum mechanics is ultimately incomplete, and that a complete theory would provide
descriptive categories to account for all observable behavior and thus avoid any indeterminism. \dots

\ldots In 1964, John Bell showed that if local hidden variables exist, certain experiments could be performed involving quantum entanglement
where the result would satisfy a Bell inequality. \dots

Physicists such as Alain Aspect \cite{Aspect:1981nv} and Paul Kwiat \cite{Kwiat:1998gc} have performed experiments that have found violations
of these inequalities up to 242 standard deviations[14] (excellent scientific certainty). This rules out local hidden variable theories.

\dots Gerard 't Hooft \cite{Hooft1,Hooft2} has disputed the validity of Bell's theorem on the basis of the superdeterminism loophole
and proposed some ideas to construct local deterministic models.``

We want to pay attention to what we talked about the associative observables. That is natural for physical quantities in quantum mechanics.
But in Ref.~\cite{Dzhunushaliev:2007vg} the possibility of consideration of nonassociative hidden variables is discussed. In this case these quantities are unobservable ones.

Let us consider what happens in our case. We have supersymmetric decomposition of (probably generalized) momentum operator \eqref{4-20}.
The constituents $Q_{a, \dot a}$ are unobservable according to the nonassociative properties \eqref{2-70a}-\eqref{2-70h} and \eqref{3-190}-\eqref{3-210}.
Following this way, we can say that we have unobservable nonassociative operators $Q_{a, \dot a}$ that are similar to hidden variables.
The main difference compared with the standard hidden variables is  unobservability of the nonassociative hidden-like variables.

\section{Discussion and conclusions}

Thus we have considered a nonassociative generalization of supersymmetry. We have shown that:
(a) one can choose such a form of 3-point associators that the corresponding Jacobiators are zero;
(b)
there is the relation between 3- and 4-point associators; (c) using these expressions, one can find 4-point associators.

We have seen that in all definitions of associators there is the Planck constant and some characteristic length $\ell_0$.
The presence of the Planck constant permits us to make natural assumptions that these associators can be regarded
as a nonassociative generalization of the Heisenberg uncertainty principle. In this case the characteristic length $\ell_0$
will be a new fundamental constant, and the corrections arising in this case have the order of $\hbar/\ell_0$. For example,
if $\ell_0 \approx \Lambda^{-1/2}$ (where $\Lambda$ is the cosmological constant) then the dimensionless corrections $\approx 10^{-120}$,
i.e., are negligible. Instead of introducing a fundamental length $\ell_0$, we can introduce a fundamental momentum $\mathcal P_0= \hbar/\ell_0$.
Physical consequences of introducing new fundamental quantities $\ell_0$ or $\mathcal P_0$ (that are consequences of nonassociativity) are:
\begin{itemize}
  \item There appears a minimum momentum $\mathcal P_0$.
  \item There appears a maximum length $\ell_0$.
  \item The appearance of the maximum length $\ell_0$ leads to the fact that the curvature is bounded below: $R_{min} \approx 1/\ell_0^2$.
  \item The minimum momentum $\mathcal P_0$ and the maximum length $\ell_0$ are connected by the Heisenberg uncertainty principle:
      $\mathcal P_0 \ell_0 \approx \hbar$.
  \item The experimental manifestation of possible nonaccosiativity can
  arise
  only for a physical phenomenon when either the momentum
      $p \approx \mathcal P_0 \approx 10^{-60} \text{kg} \cdot \text{m} \cdot \text{s}^{-1}$
      or on the scales $l \approx \ell_0 \approx \Lambda^{-1/2} \approx 10^{26}$m (if the fundamental length $\ell_0 \approx \Lambda^{-1/2}$).
\end{itemize}
We have also discussed a possible interpretation of nonassociative supersymmetric generators $Q_{a, \dot a}$
as hidden-like variables in quantum theory. The main idea here is that the nonassociativity leads to unobservability of these variables.

\section*{Acknowledgements}

This work was supported
by a grant $\Phi.0755$  in fundamental research in natural sciences by the Ministry of Education and Science of Kazakhstan. I am very grateful to V. Folomeev for fruitful discussions and comments.

\end{document}